\documentclass[10pt]{iopart}

\usepackage{graphicx}
\usepackage{wrapfig}
\usepackage{iopams}  
\begin{document}

\title{Trapped Ion Imaging with a High Numerical Aperture Spherical Mirror}

\author{G Shu, M R Dietrich, N Kurz, and B B Blinov}

\address{Department of Physics, University of Washington, Seattle, WA 98105-1560 USA}
\ead{shugang@u.washington.edu}
\begin{abstract}
Efficient collection and analysis of trapped ion qubit fluorescence is essential for robust qubit state detection in trapped ion quantum
computing schemes. We discuss simple techniques
of improving photon collection efficiency using high numerical aperture (N.A.) reflective optics. To test these techniques we placed a spherical mirror 
with an effective N.A. of about 0.9 inside a vacuum chamber in the vicinity of a linear Paul trap. We demonstrate stable and reliable trapping 
of single barium ions, in excellent agreement with our simulations of the electric field in this setup. While a large N.A. spherical mirror
introduces significant spherical aberration, the ion image quality can be greatly improved by a specially designed aspheric corrector lens located outside
the vacuum system. Our simulations show that the spherical mirror/corrector design is an easy and cost-effective way to achieve high photon collection 
rates when compared to a more sophisticated parabolic mirror setup.
\end{abstract}

\submitto{\JPB}
\maketitle

\section{Introduction}
Trapped ion quantum computing is a rapidly progressing and developing area of experimental and theoretical research~\cite{haffner2008}. 
All basic building blocks
of a quantum information processor based on trapped ions have been demonstrated: qubit initialization, typically via optical pumping; 
qubit state manipulation by RF~\cite{blinov2004} or optical~\cite{blatt2004} fields; qubit state readout by quantum
jumps~\cite{blatt1988}. Fundamental quantum logic gates~\cite{turchette1998} and multi-qubit entanglement~\cite{wineland2005,blatt2005} 
have also been demonstrated. In this paper we address the issue of qubit state readout which is typically based on the ion bright/dark
state discrimination. Efficient ion fluorescence collection is critical for fast, reliable qubit state detection required not only at the end of any
quantum computation, but throughout the computation process to enable quantum error correction~\cite{shor1995,steane1996}.  
Additionally, higher photon collection rates from single trapped ions or atoms would lead to more efficient single-photon sources
and ion-photon entanglement~\cite{maunz2007}.

Ion fluorescence collection is commonly performed using reasonably high-N.A. refractive optics, achieving a collection 
efficiency of a few percent. A major practical limitation in using this type of optics is the large working distance, of order of a few 
cm, due to the fact that
the lens is located outside the vacuum chamber. High-N.A. optics in this case becomes a very expensive, custom job. Some efforts have been made to place
the lens inside the vacuum apparatus, closer to the ion trap~\cite{eschner2001,koo2004} to achieve up to about 5\% light collection efficiency.

Another approach is to use reflective optics such as an optical cavity, or a single spherical or parabolic mirror. 
A small (order of 100 $\mu$m), high-finesse optical cavity similar to those used in cavity QED experiments~\cite{miller2005, maunz2004}
appears to be the ultimate solution. Many serious efforts are under way to couple trapped ions to a high-finesse cavity ~\cite{lange2004,Drewsen2008} and rapid progress is being made. However, the presence of dielectric cavity mirrors in the close vicinity of the trapped ions may be detrimental to the ion trapping process.
A parabolic mirror surrounding the ion trap is a attractive alternative~\cite{sondermann2007}. It would provide excellent performance in terms of image quality.
However, making a high-N.A. parabolic mirror itself is quite a demanding task. Our simulations of a parabolic mirror show that it is also
sensitive to ion's alignment with respect to the focus of the mirror. A spherical mirror, due to its high symmetry, is much less sensitive 
to such misalignment.

A spherical mirror placed in vacuum with ions in the center of the mirror
curvature has been used to double the photon count on a large area photomultiplier tube~\cite{flatt2007}, but was not used for imaging.
Here we present an ion imaging system based on 0.9 N.A. spherical mirror which is in principle capable of collecting up to about 28\% of light
emitted by the ion. 
To improve the image quality which suffers greatly from strong spherical aberration introduced by the high-N.A. mirror 
we designed an aspherical corrector lens based on 
the widely used Schmidt telescope 
corrector element. Our simulations show that with this corrector the ion image is suitable not only
for high-quality imaging and photon counting, but possibly even for fiber-coupling single ion fluorescence.

\section{Simulations of ion trap and optical system}

To model the ion trap behavior in the presence of a spherical mirror we used the method of images~\cite{griffith}. We compared the trapped ion orbit 
in an unperturbed quadrupole electric field produced by the four RF electrodes of our linear trap, described in much detail in the following section, 
to the ion motion in the presence of a grounded metallic symmetry plane nearby, which is a simplified model of the spherical metallic mirror. Our
simulations show that the ion orbit is perturbed by less than 1~$\mu$m if the plane is farther than 5~mm away. 
This displacement of the trapping center is negligible and, if necessary, can be easily 
corrected by applying a small bias voltage to the top RF electrodes. Figure~1 shows the calculated orbit displacement of a trapped ion in 
the trap in the presence of a spherical mirror.

\begin{figure}
\includegraphics{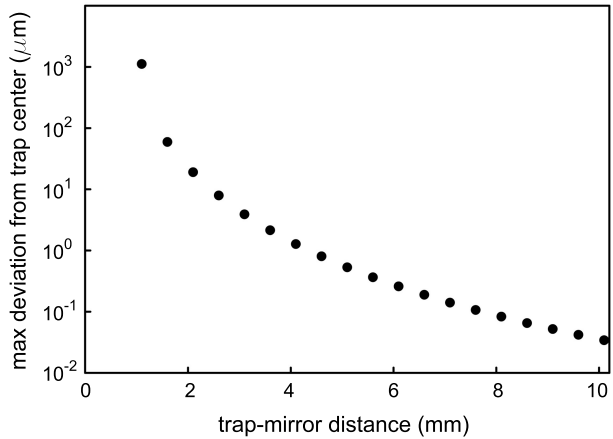}
\caption{Trapped ion orbit r.m.s. displacement from its unperturbed position due to the presence of a grounded metallic surface in the
vicinity of the trap as a function of the distance between the trap center and the surface.}
\end{figure}

Due to restrictions imposed  by our ion trap construction, the size of the reflective surface is limited to 20$\times$20~mm$^2$. 
To achieve the highest numerical aperture, the mirror should stay as close as possible to the 
trapping center. Several considerations prevent it from being too close: the mirror should not block the cooling laser access arranged at right angle
to the mirror axis; and the entirety of the mirror surface should 
be shielded from the atomic barium beam to avoid coating. Additionally, it should not significantly affect the trap stability as discussed above. 
We found a suitable spherical mirror with focal length of 10~mm in Edmund Optics catalog (stock no. NT43-464) and set out to model its performance.

\subsection{Aspheric corrector lens}
Using an aspheric plate to correct the aberration of the image by a spherical mirror was invented by Bernhard Schmidt in 1930 (see e.g.~\cite{born}). 
The concept is to distort the collimated light beam before it hits the mirror to cancel most of the spherical aberration 
in reflection. This is achieved by adding a piece of optics through which the spherical mirror looks like a parabolic mirror. 
It is widely used in telescopes and reflective camera lenses. To lowest order, the general corrector curve equation can be expressed as:
\begin{equation}
z=\frac{(r^4-k r^2)d^4}{4(n-1)R^3}
\end{equation}
where $z$ is the thickness of the corrector, $r$ is the distance from the optical axis, $R$ is the radius of curvature of the spherical mirror, 
and $n$ is the refractive index of the corrector material. 
$k$ is a constant determined by the corrector plate's position with respect to the mirror, which is normally within $R$. 
In our situation, the problem is reversed: we are attempting to create a collimated beam of light emitted by a point source (the ion)
reflected from a spherical mirror. An additional restriction is that
the viewport is about 17~mm away from the mirror focus, so $k$ should be equal to 0, corresponding to a plate at the center of the mirror curvature. 
Through this general formula, we arrive at a fourth-order curve equation for our specific parameters (using standard BK7 glass):
\begin{equation}
z=5.98785\times 10^{-5}r^4. 
\end{equation}
This curve is plotted in Figure~2(a) as the black solid line. However, the 4$^{th}$-order curve works well only for small N.A. mirrors. In
our case, higher order terms (and higher order aberrations, and aberrations due to the planar viewport) 
would have to be taken into account to determine the optimal corrector shape.
For typical systems such as a reflective telescope or a Schmidt camera, better performance 
can be achieved by adding a small lens between the focus and the mirror, but this would not work 
for us since the refractive optics placed too close to the trap would destabilize the ion by patch charging.

\begin{figure}
\includegraphics{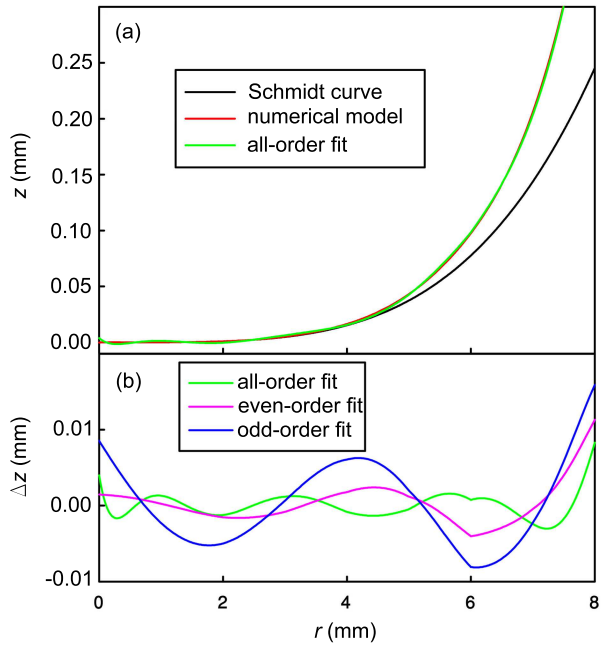}
\caption{(a) Thickness $z$ of various corrector plate shapes plotted vs. distance $r$ from the optical axis: the Schmidt corrector
(black), the numerical model curve (red) and the 10$^{th}$ order polynomial fit to the numerical result (green). (b) Deviations of various types
of polynomial fits from the exact numerical model.}
\end{figure}

As an alternative, we tried a brute-force approach by directly calculating the best correcting element shape for our specific set up. The principle 
is quite simple: ensure that any ray originating from the ion and reflected off the mirror is parallel to the optical axis. 
Another option would be to make the deflected ray slope proportional to 
its distance from the main optical axis. We are not attempting to derive a general solution here, rather the one ideal for our application. 
We used Wolfram Research, Inc. Mathematica to numerically perform the ray tracing and curve fitting. Because of the rotational symmetry, 
the problem can be reduced to 
two dimensions. The slopes of the ray span are recursively calculated at the curve vertex $z$-value and then integrated and expressed as a 
function of $r$, the distance from the optical axis. The deviation from the vertex is added to the thickness of the corrector plate for the 
next round of calculation. The curve equation converged quickly and the difference was smaller than a quarter of the wavelength after only 10 iterations. 
Again, the standard BK7 glass was used in the simulation.

This numerical simulation predicted an irregular curve shown in Figure~2(a) as a red line. 
We used the optics modeling software (Sinclair Optics, Inc. OSLO~LT) to evaluate the performance of such a corrector element. 
Since OSLO~LT only supports polynomials up to order 10 for its surface data input, it was impossible to exactly reproduce the shape
of the corrector in the simulation. 
Instead, we fit the corrector shape profile with even, odd and full 10$^{th}$ order polynomials; they differ by as much as 20$\mu$m 
(about 40 times the wavelength at 493~nm) from the numerically derived shape. 
The even-order polynomial fit is very good close to the optical axis but gets much worse than the full 
10$^{th}$ order polynomials fit in off-axis region. The odd-order polynomial has the worst fit overall. The all-order polynomial fit is plotted
in Figure~2(a) in green, and the deviations of the three types of fits from the numerical model are plotted in Figure~2(b).

Surprisingly, all of the fits perform much better than the Equation~2 curve in correcting the spherical aberration. 
We estimated the corrector's performance by modeling how well the collimated beam it produces is focused by an 
infinity-corrected microscope objective~\cite{alt2002} (Figure~3(a)). As shown in Figure~3 (b), 
both the even- and full-order curves gave a root mean square spot size of 67~$\mu$m, about 1/3 the spot size produced by the 4$^{th}$ order curve
in Equation~2. With some basic spatial filtering techniques, 
any background from light scattered from other surfaces and other sources can be largely suppressed 
so that a PMT can be used for high-speed, accurate photon detections with low noise. 
If the actual numerically-derived corrector shape is used rather than a fit, we believe the performance should be close to 
that of a perfect parabolic mirror, and nearly diffraction-limited imaging and single-mode fiber coupling should be possible. Off axis displacement up to 100$\mu$m will increase the spot size by less than 5\%, which makes it possible to resolve a string of tens of ions 5 to 10 $\mu$m apart.

\begin{figure}
\includegraphics{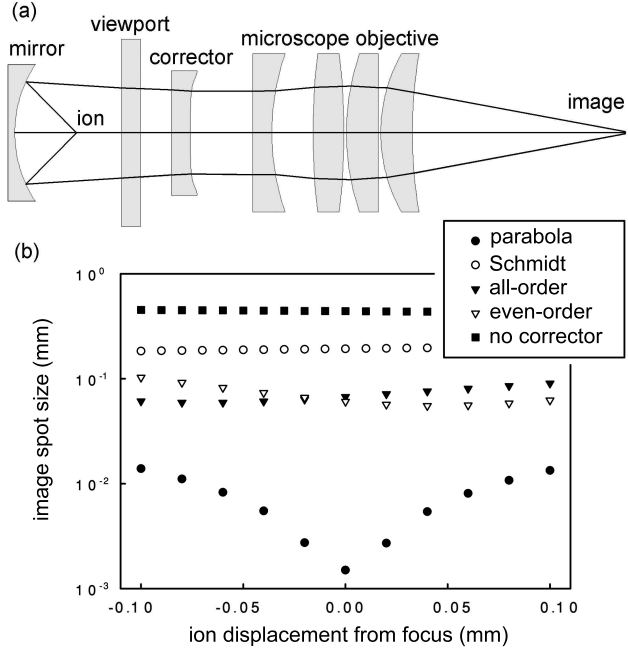}
\caption{
Simulation of the corrector performance. 
(a)
The simulation model.
 A point-source of light is reflected off the spherical mirror and, after passing through a flat viewport window and the corrector plate, is imaged with the microscope objective. To model the parabolic mirror performance the corrector element was taken out. 
Note that in the preliminary result reported in section 4, the corrector and the multi-element lens were not used.
 (b) Spot sizes produced by different shape correctors as a function of ion displacement from the focus along the optical axis. The diffraction limit for this setup at 493~nm is about 1.5~$\mu$m.}
\end{figure}

\section{Experimental setup}
To test the spherical mirror performance we upgraded our existing ``5-rod'' linear Paul trap (shown in Figure~4) 
originally designed by the Monroe group~\cite{blinov2005}. 
An aluminum frame supports the four RF quadrupole rods and the two endcap needle electrodes, all made from 0.5~mm dia. tungsten wire, 
separated from each other and from the ground by alumina spacers. The needles are electrolytically etched from the cylindrical wire stock. 
The electrodes are glued to the alumina spacers with a UHV-compatible ceramic adhesive (Scientific Instrument Services, Inc. SCC8). 
The opposite rod centers are about 1.4~mm apart, and end caps are separated by about 2~mm. 

\begin{figure}
\includegraphics{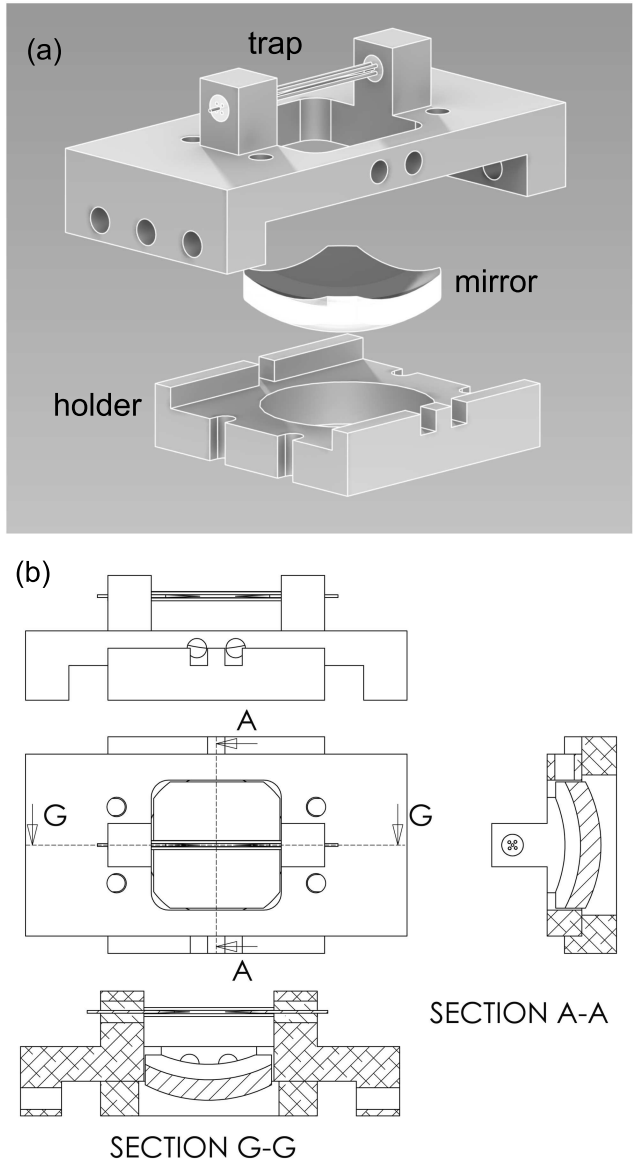}
\caption{Linear ion trap with integrated spherical mirror. (a) An exploded view of the assembly setup. The mirror holder allows precision positioning
of the mirror with respect to the trap. (b) Technical drawing showing the details of the assembly. Dimensions are available upon request.}
\end{figure}

Our mirror is a commercial-grade aluminum-coated spherical concave 20~mm radius of curvature, 25~mm dia. mirror. 
To fit it into the trap frame which has a 20\time20~mm$^2$
opening, four edges were ground off. A special aluminum holder was made to keep the mirror normal to the grinding surface to ensure square cut. 
To fix the mirror to the frame at the desired distance from the ion trap, a precision holder (see Figure 4(a))was made to support 
the mirror from underneath such that 
its curvature center is 10~mm above the needles. The holder was fixed to the frame by 4 screws and its alignment was fine-tuned in the transverse 
direction by moving along two knife edges. The alignment procedure was done under a measurement microscope to achieve better accuracy. 
The needles were placed an 
equal distance (about 1~mm) away from the geometric center. However, after being glued into place the trap assembly apparently shifted 
with respect to the mirror, 
moving the trapping center about 50~$\mu$m away from the focal point of the mirror.

The trap and the mirror were mounted inside a Kimball Physics, Inc. 4-1/2'' Conflat spherical octagon vacuum chamber with four 1.33'' Conflat 
viewports used for laser access, and one 3.1~mm thick, 1.5'' re-entrant fused silica viewport for ion imaging. To load barium ions into the trap, 
a simple oven filled with
metallic Ba was installed near the trap. The oven is made of a 1~mm diameter alumina tube sealed on one end, and the other end pointing at the trap,
with thin tungsten wire wrapped around. The oven can be easily heated up to a few hundred degrees C by running about 1~A of current through the
tungsten filament. To collimate the atomic barium beam and prevent barium coating of sensitive trap surfaces and the viewports, a 1~mm aperture was
placed between the oven and the trap. A 20 l/s ion pump and a Ti sublimation pump maintain ultrahigh vacuum (better than $10^{-10}$~torr) around the
trap. The trap is driven by an RF voltage with a few hundred volts amplitude at about 22~MHz generated by a high-Q helical 
resonator~\cite{macalpine1959} fed by about
1 watt of RF; the endcaps are kept at +100~V DC. This produces a trapping potential with the transverse secular motion frequencies 
$\omega_{x,y}/2\pi\approx 1$~MHz and the axial frequency $\omega_a/2\pi$ of about 100~kHz.

We load Ba ions into the trap by photoionizing~\cite{chapman2007} neutral barium with a deuterium lamp. The lamp has 
significant ultraviolet spectral content near 238~nm, which is the ionization threshold for neutral barium. We find that this
direct photoionization is very efficient and we are able to load ions with very low Ba oven temperatures.

Barium ions are laser-cooled on the strong $6S_{1/2}-6P_{1/2}$ transition near 493~nm, with an additional laser near 650~nm for repumping out of
the metastable $5D_{3/2}$ state. A detailed description of the laser system can be found in~\cite{kurz2008}.

\section{Preliminary results}
We first set up the imaging system in a way similar to our standard setup. A long working distance, 0.25~N.A. homebuilt 
multi-element lens 
designed after Ref.~\cite{alt2002} was used
\begin{figure}
\includegraphics{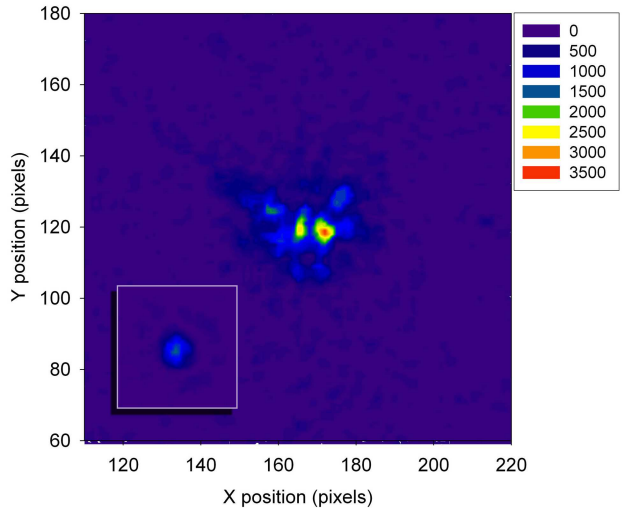}
\caption{Single Ba$^+$ ion image taken by the EMCCD camera using the 0.9~N.A. spherical mirror, compared to the image of the same ion by a 0.25~N.A. multi-element lens with total magnification of 70 (inset). Exposure time was 0.8~s for both images. The pixel size is 10~$\mu$m. Image intensity is in the (same) arbitrary units. The integrated intensity of the ion image by mirror is about 7.5 times higher than that by multi-element lens.}
\end{figure}
to form an intermediate image of the ion. A spatial filter (0.5~mm pinhole) was placed in the image plane to reduce the background light including any reflection from the mirror. The
intermediate image was then re-imaged onto an Electron-Multiplying CCD (EMCCD) camera (Andor$^{TM}$ Luca$^{EM}$) with a 25~mm focal length doublet lens.

A narrow-band, 493~nm interference filter was placed in front of the camera to reduce the background even further.
The trap was shown to work remarkably well, holding single ions for several days even without the laser cooling, in an excellent
agreement with our electric field modeling. An image of a single
laser-cooled Ba$^+$ ion generated by this system is shown in the inset of Figure~5.

In a preliminary test run, we attempted to image the ion onto an EMCCD camera with the spherical mirror alone (i.e. without using
any corrector element or the microscope objective). To our surprise we found 
a very sharp image of a single ion about 60~cm from the mirror (Figure~5). It is from this observation that we conclude that the trap center 
moved away from the mirror surface during the assembly of the trap.
The distorted image shape is 
probably due to a combination of the spherical aberration and the coma due to a misalignment of the ion location with the optical axis of the mirror. 
The observed increase in the photon count rate of a factor of about 7.5 was not as good as we expected from the relative ratio (approximately 15) of the solid angles of the 0.9~N.A. and 0.25~N.A. optics, probably because of the large spherical aberration. 
A large fraction of the light emitted by the ion is actually diffused around the bright image spot, and falls outside the photon counting aperture.
The background (which was subtracted by the camera control software) was rather high due to lack of spatial filtering.     

\section{Conclusions}
In summary, we built and successfully operated a linear Paul ion trap with an integrated spherical mirror and directly imaged and 
resolved single ions without any additional optics. A future improvement for correcting spherical aberration with an aspheric lens is proposed and 
analyzed. With these simple and inexpensive modifications we hope to increase the trapped ion fluorescence collection efficiency 
by more than an order of magnitude. 

\ack
We would like to thank Nathan Pegram for help with early parts of this work and Edan Shahar, Aaron Avril, Chris Dostert, Frank Garcia and Thomas Reicken 
for their assistance. This research was supported by the National Science Foundation Grant No. 0758025, the Army Research Office, 
and the University of Washington Royalty Research Fund.

\end{document}